\documentclass[a4paper,11pt]{article}
\usepackage{jheppub}
\usepackage{amsmath}
\usepackage{amsfonts}
\usepackage{amssymb}
\usepackage{textcomp}
\usepackage{graphicx}% Include figure files
\usepackage{bm}% bold math
\usepackage{color}
\usepackage{verbatim}
\usepackage{graphicx,color}
\usepackage{epsfig}
\usepackage{subfigure}
\usepackage{amsfonts,amsmath,amssymb}

\DeclareMathAlphabet{\mathpzc}{OT1}{pzc}{m}{it}

\begin{document}

\rightline{UTTG-19-16}
\vskip 1cm

 \title{Boundary Causality vs Hyperbolicity for Spherical Black Holes in Gauss-Bonnet}
 
 \author[a]{Tom\'as Andrade}
 \affiliation[a]{Rudolf Peierls Centre for Theoretical Physics \\ University of Oxford, 1 Keble Road, Oxford OX1 3NP, UK} 
  
\author[b,c]{Elena C\'aceres}
\affiliation[b]{Facultad de Ciencias, Universidad de Colima, Bernal Diaz del Castillo 340, Colima, Mexico}
 \affiliation[c]{Theory Group, Department of Physics, University of Texas, Austin, TX 78712, USA}

\author[d]{Cynthia Keeler}
\affiliation[d]{Niels Bohr International Academy, Niels Bohr Institute \\
University of Copenhagen, Blegdamsvej 17, DK 2100, Copenhagen, Denmark}

% e-mail addresses: one for each author, in the same order as the authors
\emailAdd{tomas.andrade@physics.ox.ac.uk}
\emailAdd{elenac@zippy.ph.utexas.edu}
\emailAdd{keeler@nbi.ku.dk}

\abstract{We explore the constraints boundary causality places on the allowable Gauss-Bonnet gravitational couplings in asymptotically AdS spaces, specifically considering spherical black hole solutions. We additionally consider the hyperbolicity properties of these solutions, positing that hyperbolicity-violating solutions are sick solutions whose causality properties provide no information about the theory they reside in. For both signs of the Gauss-Bonnet coupling, spherical black holes violate boundary causality at smaller absolute values of the coupling than planar black holes do. For negative coupling, as we tune the Gauss-Bonnet coupling away from zero, both spherical and planar black holes violate hyperbolicity before they violate boundary causality. For positive coupling, the only hyperbolicity-respecting spherical black holes which violate boundary causality do not do so appreciably far from the planar bound.  Consequently, eliminating hyperbolicity-violating solutions means the bound on Gauss-Bonnet couplings from the boundary causality of spherical black holes is no tighter than that from planar black holes.}

 \maketitle

%%%%%%%%%%%%%%%%%%%%%%%%%%%%%%%%%%%%%%%%%%%%%%%%%%%%%%%%%%%%%%%%%%%%%%%%%%%%%%%%%%%%% 
\section{Introduction}
\label{sec:intro}
%%%%%%%%%%%%%%%%%%%%%%%%%%%%%%%%%%%%%%%%%%%%%%%%%%%%%%%%%%%%%%%%%%%%%%%%%%%%%%%%%%%%% 

When considering higher-curvature corrections to general relativity, there are two ways to restrict the parameter space of corrections.  First is the top-down approach, starting from a consistent string theory and deriving the specific tower of curvature corrections it produces. Second, one can also investigate the possible curvature corrections by a bottom-up approach, restricting the parameter ranges by requiring, for example, causal consistency, or perhaps hyperbolicity of the equations of motion.  This paper will follow the second approach, seeking to understand limitations on possible Gauss-Bonnet (GB) couplings.

In order for a theory in Anti-de Sitter (AdS) space to have a consistent dual CFT, the AdS bulk should not propagate information faster than the boundary theory. The papers \cite{Brigante:2008gz,Buchel:2009tt,Buchel:2009sk,Camanho:2009hu,Camanho:2009vw,Camanho:2010ru} all study the restrictions this notion of  \emph{boundary causality} places on GB curvature corrections. For a theory with GB coupling outside of the window $- 7/36 < \lambda < 9/100 $, planar black holes in AdS space allow modes to propagate through the bulk faster than along the boundary.  This is possible because gravitational modes propagate along characteristic surfaces of the PDEs, rather than along light cones.

More recently \cite{Camanho:2014apa} considered limits placed on higher-curvature couplings due to graviton three-point amplitudes, concluding that causality issues (and in fact closed timelike curves) will be present unless there is an infinite tower of higher-curvature corrections.  However, their most detailed calculations for AdS spaces involve the singular shockwave (Aichelburg-Sexl) metric; as pointed out by \cite{Papallo:2015rna}, perhaps we should throw out these specific solutions rather than disallowing non-zero GB couplings.  As partial evidence that the AS metrics are sick solutions, \cite{Papallo:2015rna} study their hyperbolicity properties.

\emph{Hyperbolicity} here refers to well-posedness of the initial value problem. AdS itself of course violates \emph{global} hyperbolicity, because data on a spacelike slice does not alone provide sufficient information to find the field profiles at all future times; one must also add boundary conditions at spatial infinity (in AdS/CFT this extra data can be thought of as choosing which sources are turned on in the CFT).  However even in AdS the equations of general relativity are locally hyperbolic; data on a time slice allows us to find the solution throughout the causal development of that slice.

Classical general relativity will always be hyperbolic in this local sense. Adding higher curvature corrections can however cause local violations of hyperbolicity.  Initial data on a particular time slice may not be propagatable under the (classical) equations of motion from curvature corrections; this is the complaint \cite{Papallo:2015rna} register against the AS spacetimes studied in \cite{Camanho:2014apa}.  In fact in some spacetimes the situation can be even worse; there may exist \emph{no} time slice providing good initial data even in a local sense; this is the hyperbolicity violation studied in \cite{Reall:2014pwa} and the one we focus on here.

There are several large questions we wish to address: Can we restrict the parameter space of higher curvature theories via a bottom-up method, or is it necessary to include all higher curvature couplings in order to have a causally consistent theory?  When we find a solution with unsavory causal properties, should we throw out the solution or the theory it lives within? 

In order to begin to answer these questions, we focus on a simple variant: what do black holes outside of the planar limit, in AdS spacetimes, teach us about GB theories?  The boundary of AdS gives a clean, easily testable definition of causality, and the work of \cite{Reall:2014pwa,Papallo:2015rna} provides both a set of effective metrics for the propagation of all graviton modes, as well as a simple test for local hyperbolicity violation.

We find that spherical black hole geometries in   GB theories are more strongly 
constrained by boundary causality considerations than their planar analogues, in the sense that the range of allowed values of the 
GB coupling decreases with the radius of the black hole. Interestingly, the restrictions that arise by demanding hyperbolicity 
of such geometries are generically stronger than the ones coming from boundary causality. For completeness, we include in our 
analysis the results of \cite{Cai:2001dz, Cho:2002hq}, which show that thermodynamic stability places further constraints
on the space of parameters of GB black holes. 

This paper is organized as follows: in section \ref{sec:hyperbolicity} we review some general results regarding causality and hyperbolicity in higher curvature theories of gravity. We later specialize these results to the case of spherical GB black holes in section \ref{sec:GB hyperb},
where we also describe in detail our method to detect boundary causality violations in spherical black hole geometries. We summarize and discuss our results, and present some interesting future directions of our work in section \ref{sec:discussion}.

%%%%%%%%%%%%%%%%%%%%%%%%%%%%%%%%%%%%%%%%%%%%%%%%%%%%%%%%%%%%%%%%%%%%%%%%%%%%%%%%%%%%% 
%\section{Causality in Higher Derivative theories}
%\label{sec:causality}
%%%%%%%%%%%%%%%%%%%%%%%%%%%%%%%%%%%%%%%%%%%%%%%%%%%%%%%%%%%%%%%%%%%%%%%%%%%%%%%%%%%%% 

%%%%%%%%%%%%%%%%%%%%%%%%%%%%%%%%%%%%%%%%%%%%%%%%%%%%%%%%%%%%%%%%%%%%%%%%%%%%%%%%%%%%% 
\section{Causal structure and hyperbolicity in higher curvature theories}
\label{sec:hyperbolicity}
%%%%%%%%%%%%%%%%%%%%%%%%%%%%%%%%%%%%%%%%%%%%%%%%%%%%%%%%%%%%%%%%%%%%%%%%%%%%%%%%%%%%% 

In this section we will review the basics of  hyperbolicity in higher curvature theories closely following 
\cite{Reall:2014pwa, Izumi:2014loa}.
%
%and  some results related to boundary causality violation in large Gauss-Bonnet black holes 
%
%
%\cite{Brigante:2007nu,Brigante:2008gz,Buchel:%2009sk,Camanho:2009hu}.
%
%A powerful method to analyze the causal properties of a system of partial differential equations is the method of characteristics. A characteristic hypersurface is one beyond which the evolution of the differential  equations is not unique; we always want to pick initial data on a (complete) non-characteristic surface so we can propagate the data consistently and uniquely. 

A powerful method to analyze the causal properties of a system of partial differential equations is the method of characteristics, which 
is based on identifying the characteristic hypersurfaces defined by these evolution equations. 
A characteristic hypersurface is one beyond which the evolution of the differential equations is not unique. Usually, we want to choose 
initial data on a complete, non-characteristic (Cauchy) slice and evolve this initial data in time\footnote{Alternatively, we can 
consider the evolution problem defined on the characteristic surfaces themselves. This approach has proven to be  fruitful 
in the study of dynamical evolution of AdS gravity \cite{Chesler:2008hg, Chesler:2013lia}.}.
In Einstein gravity, the characteristic surfaces are the null surfaces of the spacetime, so as long as we start on some spacelike surface, we can evolve the initial data uniquely.  This is not the case in a theory where the characteristic hypersurfaces for some degrees of freedom may be spacelike with respect to the propagation of light rays. In \cite{Reall:2014pwa, Izumi:2014loa} the authors studied the characteristic hypersurfaces for tensor, vector, and scalar graviton degrees of freedom in spherically symmetric black hole solutions of Lovelock theories. Here we will briefly summarize the main points of these references 
which will be used in later sections.

Consider a $d+1$ dimensional spacetime with coordinates $x^\mu$ with metric which we denote as $g_I$. In Lovelock theories the equations of motion $\mathcal{E}_J=0$ depend linearly on $\partial_0^2 g_I$. Thus $\mathcal{E}_J$ can be written as
\begin{equation}\label{eq:eom}
\frac{\partial \mathcal{E}_J}{\partial(\partial^2_0 g_{I})} \partial_0^2 g_{I} + \dots =0.
\end{equation}
If we know $g_{I}$ and its derivatives in a given hypersurface $\Sigma$ with coordinates $(x_0=0, x_i)$ then, just by acting with $\partial_i$, we also know $\partial _i\partial_\mu g_I$ . However, $\partial_0^2 g_{I} $ has to be determined from the equations of motion \eqref{eq:eom}, which will only have unique solutions if the matrix
$$\frac{\partial \mathcal{E}_J}{\partial(\partial^2_0 g_{I})}, $$
also called the principal symbol, is invertible. In this case $\Sigma$ is non-characteristic and we can uniquely evolve the equations of motion starting with initial data on $\Sigma$. If the principal symbol is not invertible, $\Sigma$ is characteristic, the equations are not hyperbolic and the initial value problem with $\Sigma$ as the starting surface is ill-posed. 

\begin{figure}[h]
 \centering
 \includegraphics[width=.7\textwidth]{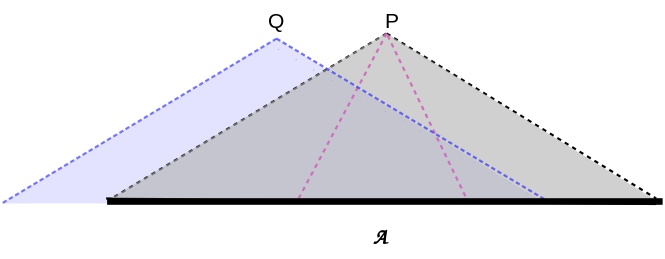}
 \caption{The causal cone is wider than the standard light cone.} \label{fig:cones}
 \end{figure}

If we consider metric fluctuations, it can be shown that the fastest mode propagates along a characteristic hypersurface. To see this consider solving the equations of motion imposing initial conditions on a  hypersurface $\mathpzc{A}$, Fig \ref{fig:cones}. The causal past of a point $P$ is given by all propagations, including any superluminal modes. The physics in $P$ is uniquely determined by the initial conditions in $\mathpzc{A}$. 

Thus the characteristic hypersurface gives the edge of the Cauchy development. 
A familiar example is Einstein gravity; characteristic hypersurfaces in GR are always null and thus gravity travels at the speed of light. 
This is not the case  in GB gravity\cite{Reall:2014pwa, Izumi:2014loa}. However,  in symmetric spacetimes, we  can define  an effective metric  such that the characteristic hypersurface for a given degree of freedom is null with respect to that metric.  

In black hole solutions of the form
\begin{equation}\label{eq:physical}
ds^2= -f(r) dt^2 +\frac{dr^2}{f(r)} + r^2 d\Omega_{d-2}^2,
\end{equation}
where $d\Omega^2_{d-2}$ is the line element of a  space $S$, linear perturbations can be classified as scalar, vector or tensor with respect to the symmetries of $S$. For each type of perturbation the equations of motion will lead to a master equation that can be written as a wave equation with a potential
\begin{equation}\label{eq:master}
\left ( -\frac{\partial^2}{\partial t^2} + \frac{\partial^2}{\partial r_*^2} - V^\ell_{A}(r)\right ) \Psi^\ell_{A}(t,r)=0
\end{equation}
where $A$ denotes the type of fluctuation, $A \in { T,V,S}$ and $\ell$ labels the harmonic.
To determine the principal symbol we need to identify  terms that involve second derivatives. We can do this by focussing on highly oscillatory modes since in this case the second derivatives will dominate the equation. For large $\ell$ and denoting as $D^2$ the Laplacian on $S$,  we can recast \eqref{eq:master} as
\begin{equation}
\left( -\frac{\partial^2}{\partial t^2} + \frac{\partial^2}{\partial r_*^2} - f(r)\frac{c_A (r)}{r^2} D^2 \right ) \Psi_A(t,r)=f(r) G_A^{\mu\nu} \partial_\mu\partial_\nu \Psi_A=0.
\end{equation}

Thus, for each mode the  characteristic hypersurface is null with respect to the corresponding effective metric:
\begin{equation}\label{eq:effective}
G_{A \mu\nu} dx^\mu dx^\nu = - f(r) dt^2 + \frac{dr^2}{f(r)} + \frac{r^2}{c_A (r)} d\Omega^2.
\end{equation}
The null cones of $G_{A \mu \nu} $ determine causality of the theory in the physical spacetime \eqref{eq:physical}.
Note that \cite{Brigante:2008gz} the factor $\frac{r^2}{f(r) c_A (r)} $ can be interpreted as the local speed of the perturbation on a constant $r$ hypersurface.

In a Lovelock theory the characteristic determinant factorizes as a product of the effective metrics
 \begin{equation}\label{eq:factorQ}
 Q(x,\xi)=( G_S^{ a b} (x)\xi_a \xi_b)^{n_S}( G_V^{ a b}(x)\xi_a \xi_b)^{n_V} ( G_T^{a b}(x)\xi_a \xi_b)^{n_T},
 \end{equation}
 where $n_S, n_V, n_T$ are the numbers of degrees of freedom of the corresponding perturbation. From \eqref{eq:factorQ} we see that if any of the $c_A$ vanishes for some $r>r_H$, then there is no surface providing good initial data for that solution; that is, the solution is not hyperbolic. 

We will see in section \eqref{sec:gb_bh} that in GB theory with a positive coupling we have $c_T>c_V>c_S$ for all $r$ and thus the tensor modes travel fastest. 
However,  $c_S$ is the only one that can become negative, so it is the scalar mode that determines if the equations of motion are hyperbolic around a given solution.
Conversely, if the GB coupling is negative, the fastest modes are the scalars and it is the tensors which are related to hyperbolicity violations.

%%%%%%%%%%%%%%%%%%%%%
%\subsection{Boundary causality in large Gauss-Bonnet Black holes}
%\EC{review of results of Myers et al  for large bh}
%\subsection{Review of Aichelburg-Sexl  metric results and small boosted black holes}
%\EC{review of Maldacena et al vs Reall \& Papallo}
%move to section on effective metrics.
%
%Recently, Reall et al studied causality in higher derivative theories. What really controls information propagation 
%are the characteristic surfaces of the PDE's, which in highly symmetric space-times admit a description in terms of 
%effective metrics -- for which the characteristic surfaces are null surfaces. \\
%Because we work in an AdS case with a GB coupling, we have three important parameters
%More recently 
% 
%%%%%%%%%%%%%%%%%%%%%%%%%%%%%%%%%%%%%%%%%%%%%%%%%%%%%%%%%%%%%%%%%%%%%%%%%%%%%%%%%%%%% 
\section{Hyperbolicity and boundary causality in spherical Gauss-Bonnet black holes}
\label{sec:GB hyperb}
%%%%%%%%%%%%%%%%%%%%%%%%%%%%%%%%%%%%%%%%%%%%%%%%%%%%%%%%%%%%%%%%%%%%%%%%%%%%%%%%%%%%% 

In a theory with superluminal modes, like GB, the causal structure is not determined by null curves but instead by the fastest modes. Thus, to analyze bulk causality in these spaces we need  information about the propagation of fluctuations in addition to the metric.  Since our goal is to elucidate the connection between bulk causality, hyperbolicity of the equations of motion, and boundary causality, we will first set up a framework that will allow us to work with null curves in the effective metric instead of with the fluctuations in the physical metric. We determine the region of parameter space for spherical black holes allowed by hyperbolicity, and  then proceed to investigate boundary causality in those backgrounds.

%%%%%%%%%%%%%%%%%%%%%%%%%%%%%%%%%%%%%%%%%%%%%%%%%%%%%%%%%%%%%%%%%%%%%%%%%%%%%%%%%%%%% 
\subsection{Effective metrics in the Gauss-Bonnet black hole}
\label{sec:gb_bh}
%%%%%%%%%%%%%%%%%%%%%%%%%%%%%%%%%%%%%%%%%%%%%%%%%%%%%%%%%%%%%%%%%%%%%%%%%%%%%%%%%%%%% 

For concreteness, we will consider a five-dimensional AdS GB black hole with spherical topology, the line element of which can be written 
as \cite{Cai:2001dz}
\begin{equation}\label{ds2_GB}
	ds^2 = - \frac{f(r)}{f_\infty} dt^2 +  \frac{dr^2}{f(r)} + r^2 d \Omega_3^2,
\end{equation}
\noindent where
\begin{equation}\label{def f}
	f(r) = r^2 \left[ \frac{L^2}{r^2} + \frac{1}{2 \lambda} \left( 1 - \sqrt{1 - 4 \lambda + 4 \lambda \frac{\mu}{r^4}}   \right)    \right].
\end{equation}
Here $\lambda$ is the GB coupling, $L$ is the AdS radius, $\mu$ a parameter related to the total energy of the black hole
and 
\begin{equation}
f_\infty=\frac{1-\sqrt{1- 4 \lambda}}{2\lambda}
\end{equation}
i%
s introduced via a trivial rescaling of the time coordinate in order to make the boundary speed of light equal to $1$. 
Note that reality of the metric requires $\lambda \leq 1/4$, the inequality being saturated for the Schwarzschild AdS black hole. 
Therefore, we restrict ourselves to the range $\lambda < 1/4.$  
%Addtionally, in this paper we only study nonnegative $\lambda$, so we have $0\leq \lambda<1/4$.
%
The horizon radius $r_H$ is the largest positive root of $f(r_H) = 0$, and it is related to the mass parameter $\mu$ by
\begin{equation}\label{def mu}
	\mu = r_H^4 + r_H^2 L^2 + \lambda L^4 
\end{equation}
\noindent provided\footnote{To see this, just plug \eqref{def mu} in \eqref{def f} and note that $f(r_H) = 0$ only if \eqref{restric lambda} 
is satisfied.  If we had chosen the opposite sign for the square root in \eqref{def f}, then smaller $r_H$ would be accessible; however, this solution is nonphysical due to ghosts. It additionally does not smoothly match onto the $\lambda=0$ solution.  Further details can be found in \cite{Cai:2001dz}.} 
\begin{equation}\label{restric lambda}
	r_H^2 + 2 \lambda L^2 > 0.
\end{equation}
This relation is automatically satisfied for positive $\lambda$ but does impose a constraint  on the  negative  values of $\lambda$ allowed for a given $r_H$. Following \cite{Buchel:2009sk}, we have chosen to parametrize the time coordinate in such a way that the boundary metric is conformal to that of Einstein's static universe 
\begin{equation}
 	ds^2 \to r^2 (- dt^2 + d \Omega_3^2 ).
 \end{equation} 
The metric above satisfies the equations of motion derived from the action 
\begin{equation}
	S = \int \sqrt{- g} \left( R + \frac{12}{L^2} + \frac{\lambda L^2}{2} ( R_{\mu \nu \rho \sigma} R_{\mu \nu \rho \sigma} - 
	4 R_{\mu \nu} R^{\mu \nu} + R^2  ) \right).
\end{equation}
From now on we will set $L = 1.$
%\TA{unless any of you is opposed? I got lazy with factors of $L$ and I don't think we gain much
%by keeping them} \\

As emphasized in \cite{Reall:2014pwa}, the causal properties of this space-time are not encoded in the geometry \eqref{ds2_GB}, 
but in the characteristic surfaces associated to the PDE's that govern the dynamics of the system. 
Moreover, the authors of \cite{Reall:2014pwa} showed that these characteristic surfaces can be of three different types: 
tensor, vector, and scalar, depending on the polarization of the gravitons that propagate on them. Moreover, the characteristic surfaces
are in one to one correspondence with the null cones of three different effective metrics, which can be written as
\begin{equation}\label{ds2_GB_eff}
	ds^2 = - \frac{f(r)}{f_\infty} dt^2 +  \frac{dr^2}{f(r)} + \frac{r^2}{c_A(r)} d \Omega_3^2,
\end{equation}
\noindent with $f(r)$ given in \eqref{def f} and $c_A$ distinguished among the different types of effective 
metric, $A = T, V, S$ where $T$, $V$, $S$ stand for tensor, vector and scalar respectively. The functions $c_A$ are given 
by
\begin{align}
 	c_T(r) & = -2 A(r) + 3 \\
 	c_V(r) &= A(r) \\
 	c_S(r) &= 2 A(r) - 1,
\end{align} 
\noindent where 
%
%\begin{equation}
%	A(r) = \frac{\gamma r^4}{1 + \gamma r^4} 
%\end{equation}
%
%\noindent with 
%
%\begin{equation}
%	\gamma = \frac{1 - 2 \lambda}{2 \lambda \mu}
%\end{equation}
%
\begin{equation}
	A(r) = \frac{(1- 2 \lambda) r^4}{ 2 \lambda \mu  + (1- 2 \lambda) r^4} .
\end{equation}

\begin{figure}[htb]
\center
\subfigure[][]{
\label{fig:allowed_hyper}
\includegraphics[width=0.39\linewidth]{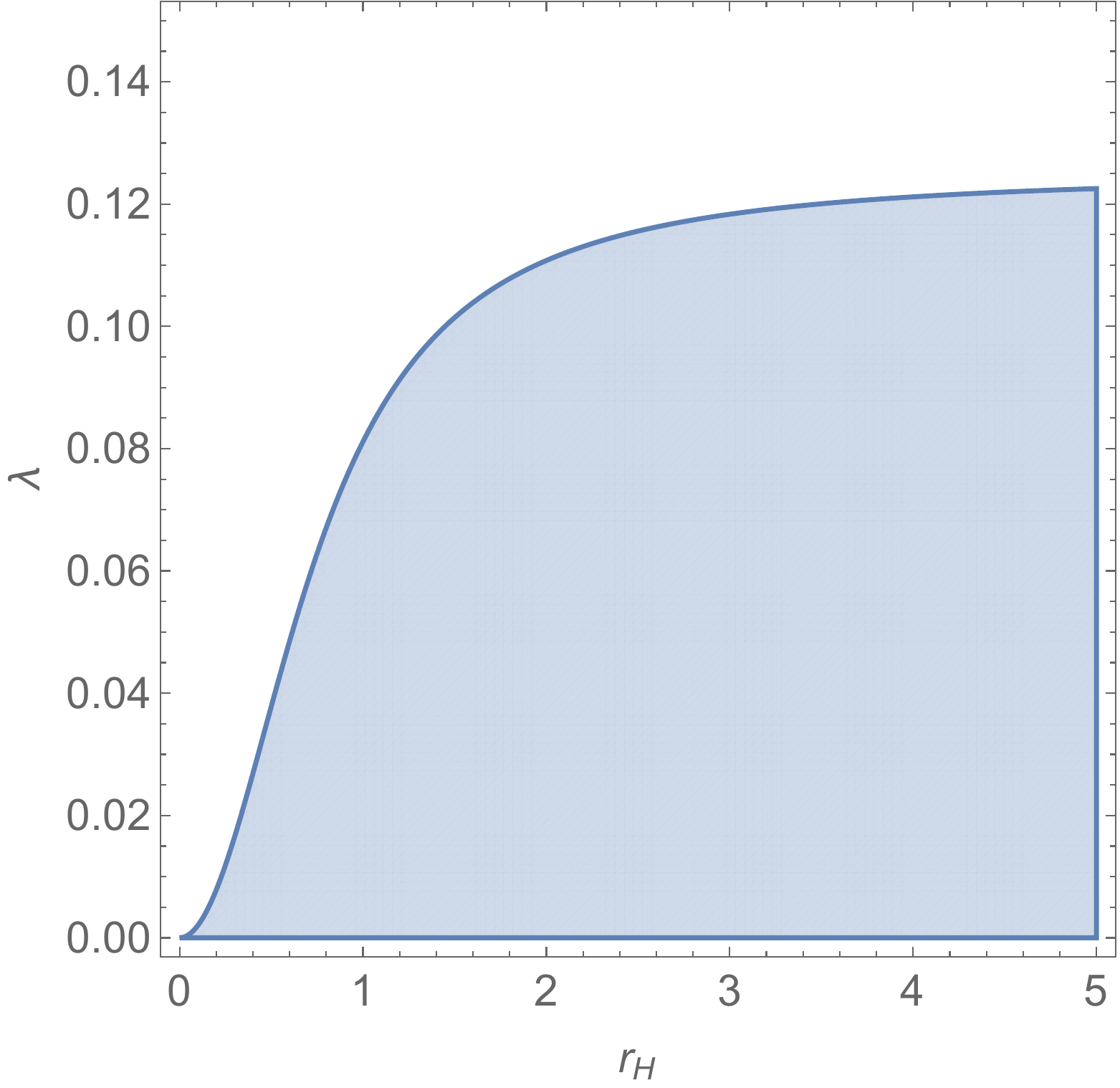}
}\qquad\qquad
\subfigure[][]{
\label{fig:allowed_cT}
\includegraphics[width=0.4\linewidth]{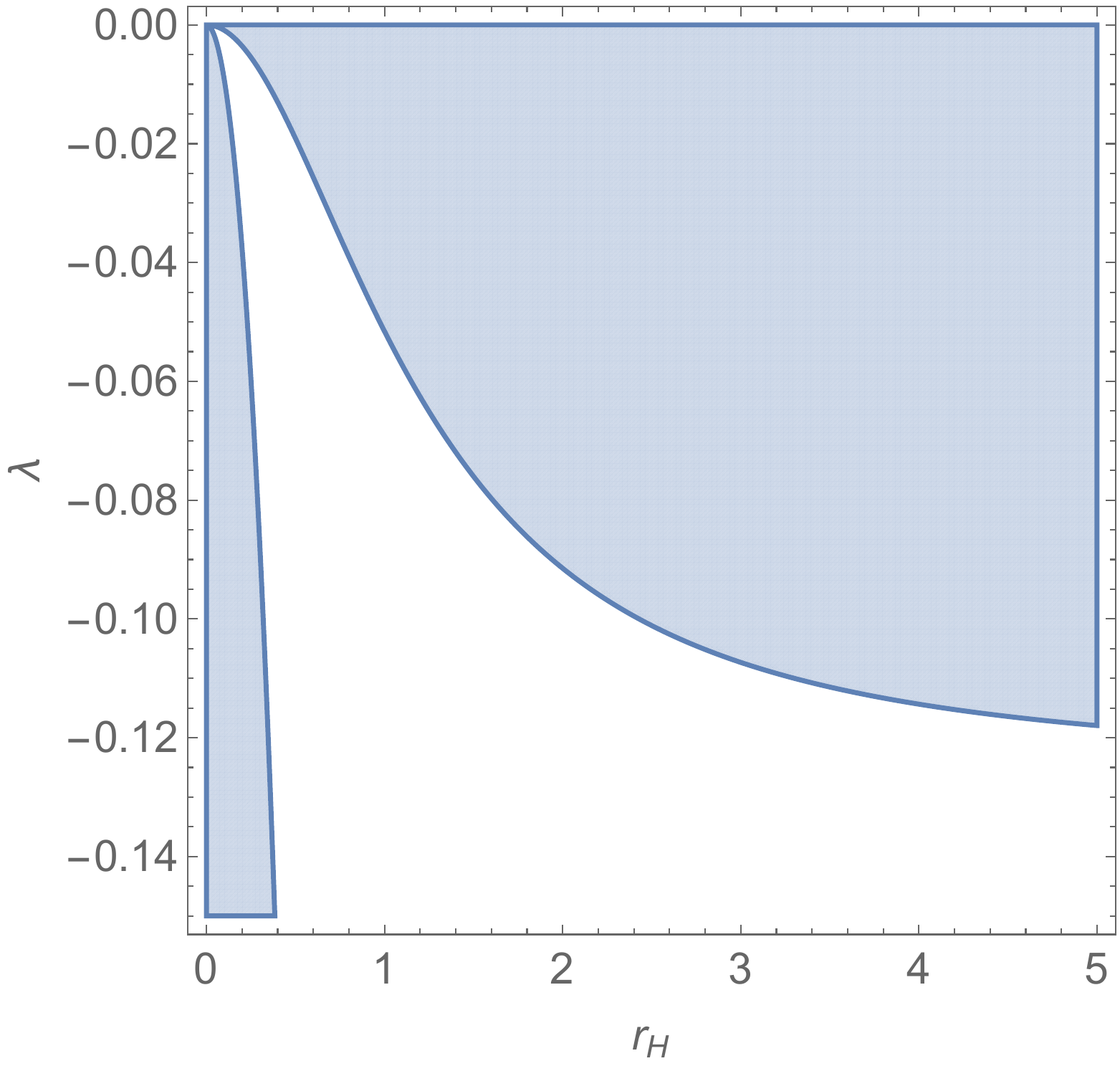}}
\caption{(Left): for $\lambda>0$, hyperbolicity  is respected if $c_S(r)>0,\, \forall r>r_H$, shaded region. 
(Right): for $\lambda <0 $ hyperbolicity  is respected if  $c_T(r)>0,\, \forall r>r_H$, shaded region. In this case the constraint $r_H^2 + 2 \lambda >0$ is not trivial and imposing it discards the vertical region.}
\end{figure}

The main result of \cite{Reall:2014pwa} is to show that some GB black hole space-times suffer from lack of hyperbolicity. 
As reviewed in section \ref{sec:hyperbolicity} in this language hyperbolicity breakdown  corresponds to any of the $c_A$ having a zero for some $r$ outside the horizon, $r_H < r < \infty$.
For  $\lambda > 0$, we can easily see that  $c_T(r) > c_V(r) > c_S(r)$ and $c_T(r) > c_V(r) > 0$, for all $r$. 
Moreover, we note that the $c_A$'s are monotonic and that by construction, $c_A\rightarrow 1$ as $r\rightarrow \infty$. Then, 
it follows that $c_S$ does not have a zero in the region $r>r_H$ if and only if $c_S(r_H) > 0$,  or, equivalently
\begin{equation}\label{hyperb bound}
	r_H^4 (1 - 8 \lambda) - 4 r_H^2 \lambda - 4 \lambda^2 \geq 0. 
\end{equation}
\noindent Black hole space-times of the form \eqref{ds2_GB}, \eqref{def f} for which \eqref{hyperb bound} is satisfied
respect hyperbolicity, see Fig. \ref{fig:allowed_hyper}. Note that in this case all effective metrics are Lorentzian outside the horizon.

We observe from \eqref{hyperb bound} that large values of $\lambda$ disfavour hyperbolicity. In fact, it is easy to see that for $\lambda > 1/8$, spherical black holes violate hyperbolicity regardless of their size. 
%while for $\lambda < 1/8$ the hyperbolicity condition is respected if 
%
%\begin{equation}
%	r_H^2 > \frac{2 \lambda}{1 - 8 \lambda} \left( 1 - \sqrt{2(1 - 4 \lambda)} \right) 
%\end{equation}
%
%This is consistent with the results of \cite{Reall:2014pwa} which found that, generically, ``small" black holes 
%are hyperbolicity violating. \\

For $\lambda<0$ the roles of the scalar, vector, and tensor modes are different. In this case $c_S(r) > c_V(r) > c_T(r)$, and only $c_T$ can have a positive root larger that $r_H$. Thus, here the tensor modes determine the region of parameter space  where the theory is hyperbolic. By an argument similar to the one above, it is easy to see that for $\lambda <0$, hyperbolicity is respected if and only if
\begin{equation}\label{hyperb bound lambda neg}
	r_H^4 (1 + 8 \lambda) + 12 r_H^2 \lambda + 12 \lambda^2 \geq 0. 
\end{equation}
\noindent See Fig. \ref{fig:allowed_cT} for a plot of this region. Note that for $\lambda < - 1/8$, all black holes violate hyperbolicity. 

As argued in the previous section, the different polarization modes (S,T, V) travel along null trajectories in the corresponding effective metric. From \eqref{ds2_GB_eff} we see that  $f(r)c_A(r)/r^2$  can be interpreted as the local speed of the graviton on a constant $r$ hypersurface
\cite{Brigante:2007nu, Brigante:2008gz}.
Since  for positive $\lambda $ we have  $c_T > c_V > c_S$ for all $r$, the tensor modes propagate fastest and determine the causal structure of the spacetime. Similarly, for $\lambda <0$, we have $c_S>c_V>c_T$; in this case the scalar modes are the fastest. Instead of following tensor or scalar  perturbations in the physical metric \eqref{ds2_GB} we can study null geodesics in the effective corresponding  metric \eqref{ds2_GB_eff}. 

%\begin{figure}
%\centering
% \includegraphics[width=.4\linewidth]{fig/allowed_hyper_draft}
%  \caption{$\lambda>0$ :  hyperbolicity  is respected if
%  %$	r_H^2 > \frac{2 \lambda}{1 - 8 \lambda} ( 1 - \sqrt{2(1 - 4 \lambda)} ) $,
%$c_S(r)>0,\, \forall r>r_H$, shaded region.}\label{fig:allowed_hyper}
%\end{figure}
 
%\begin{figure}
%\centering
%  \includegraphics[width=.4\linewidth]{fig/allowed_cT_v2}
%  \caption{ $\lambda <0 : $ hyperbolicity  is respected if  $c_T(r)>0,\, \forall r>r_H$. In this case the constraint $r_H^2 + 2 \lambda >0$ is not %trivial and imposing it discards the vertical region.}\label{fig:allowed_cT}
% \end{figure}

\subsection{Geodesic equations}

In order to study violations of boundary causality we will study geodesics which leave the boundary and then return back to it, in the effective metrics which control the propagation of the fastest modes.  Since we are working with spherically symmetric metrics, we write
\begin{equation}
d\Omega_3^2 = d \phi^2 + \sin^2\phi d\Omega_2^2,
\end{equation}
where $d\Omega_2^2$ is the metric for the unit $2$-sphere.  Below we will use spherical symmetry to study only geodesics which are constant in the $\Omega_2$ directions. 

Because we have chosen units in which the boundary speed of light is one, we expect that the presence of a boundary-anchored effective null geodesic subtending an angle $\Delta \phi$ larger than the corresponding time interval $\Delta t$ between its two endpoints will imply violation of boundary causality\footnote{This condition requires a refinement due to the compactness of the boundary and will be made precise below.}. With this intuition in mind, let us proceed to study the propagation of null geodesics in the effective spacetimes. 

The null geodesic equations in the effective metric \eqref{ds2_GB} become 
\begin{equation}
\dot{t} = \frac{f_\infty}{f(r)}, \qquad \dot{\phi} =\frac{\ell c_A(r)}{r^2}, \qquad \dot{r} = \eta \sqrt{f_\infty-\ell^2 c_A(r) f(r)/r^2}.
\end{equation}
Here $\ell$ is the angular momentum (conserved quantity due to the $\partial_\phi$ killing vector), and $\eta = \pm 1$ indicates whether we have an ingoing/outgoing geodesic.  The dot, $\cdot$, denotes the derivative with respect to the affine parameter (scaled so that the conserved quantity due to the $\partial_t$ killing vector is $1$).
Due to the $\phi \rightarrow -\phi$ symmetry, we need only study $\ell>0$.  We are looking for geodesics which both start and end at the boundary $r=\infty$, so these paths must have $\dot{r}$ real at the boundary $r=\infty$. Hence, it suffices to study the range $0<\ell<1$.
A similar calculation for AdS Schwarzschild black holes can be found in \cite{Hubeny:2013gba}. 

Since the geodesics in question must also return to the boundary, they must have a turning point in $r$.  Turning points occur when $\dot{r}=0$, and represent the minimum radius for a particular geodesic (which is a function of the $\ell$ for that geodesic).  This minimum radius $r_m$ solves
\begin{equation}\label{eq:def_rm}
f_\infty=\ell^2 c_A(r_m) f(r_m)/r_m^2.
\end{equation}

The time $\Delta t$ that a geodesic takes to travel from the boundary to its minimum radius back to the boundary is given by
\begin{equation}\label{eq:delta_t}
\Delta t = 2\int_{r_m}^\infty dr \frac{f_\infty}{f(r)\sqrt{f_\infty-\ell^2 c_A(r) f(r)/r^2}}.
\end{equation}
Here the $2$ arises because the geodesic must go in to the minimum, and then return back to the boundary. Similarly the angle $\Delta \phi$ subtended by the geodesic in this process becomes
\begin{equation}\label{eq:delta_phi}
\Delta \phi = 2\int_{r_m}^\infty dr \frac{\ell c_A(r)}{r^2 \sqrt{f_\infty-\ell^2 c_A(r) f(r)/r^2}}.
\end{equation}
For sake of completeness let us briefly   comment on geodesics without a turning point. These type of geodesics end in the singularity  and are not relevant for boundary causality. However, they enter in the definition of  holographic objects like  the causal wedge and causal holographic information. The minimum $\ell_{min}(r_h,\lambda)$ for a geodesic to have a turning point, {\it i.e.}  for \eqref{eq:def_rm} to have a solution $r_m>r_h$, can be determined numerically and geodesics without a turning point can be easily studied in our setup. 

In appendix \ref{sec:app1}, we analytically obtain an expansion for the expressions \eqref{eq:delta_t} and \eqref{eq:delta_phi}, in the limit of large minimum radius (which includes the planar black hole limit).

%%%%%%%%%%%%%%%%%%%%%%%%%%%%%%%%%%%%%%%%%%%%%%%%%%%%%%%%%%%%%%%%%%%%%%%%%%%%%%%%%%%%% 
\subsection{Method to test boundary causality}
\label{sec:results}
%%%%%%%%%%%%%%%%%%%%%%%%%%%%%%%%%%%%%%%%%%%%%%%%%%%%%%%%%%%%%%%%%%%%%%%%%%%%%%%%%%%%% 

The setup developed above is  applicable to any of the modes. However, as previously explained,  violations of boundary  causality are determined by the fastest mode, while violations of hyperbolicity are related to the mode that has a zero for some $r>r_H$.  Before presenting our results let us recall that,  

$\bullet$ For  $\lambda>0$, we have   $ c_T > c_V > c_S$ for all r. Thus,  the fastest modes  are  the tensors and to study boundary causality we will follow null geodesics in the tensor effective metric. On the other hand, in this case only $c_S$  can have a zero. Therefore, the region of parameter space where the theory is hyperbolic is determined by $c_S$.

$\bullet$ For $\lambda<0$   the fastest modes  are  the scalars, $c_S>c_V>c_T$,  and we investigate boundary causality by following  null geodesics in the scalar effective metric. The region of parameter space where the theory is hyperbolic is determined by $c_T$ .

\begin{figure}[h]
\centering
  \includegraphics[width=.4\linewidth]{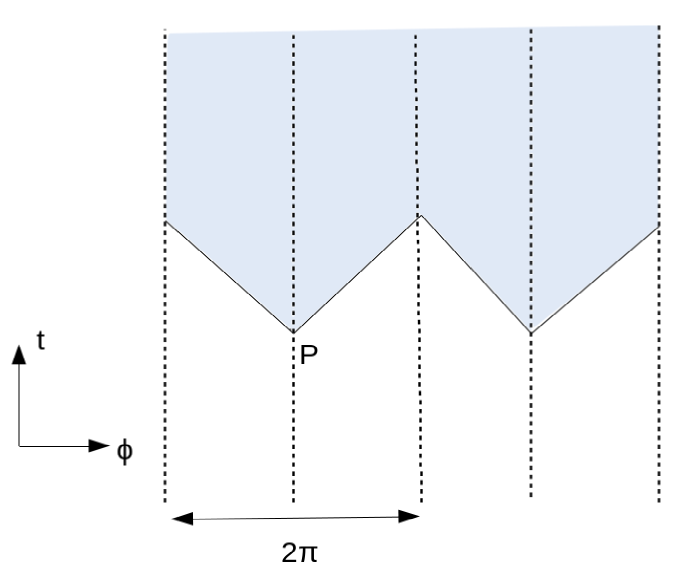}
  \caption{Null geodesic starting from boundary point $P$ does not violate causality if it returns to the boundary landing in blue region.}
  \label{fig:caus_conditions}
\end{figure}

Our task is then  to follow null geodesics in the appropriate effective metric emanating from some point $p$ at the boundary and coming back to the boundary. If we denote $\Delta \phi$ the angle subtended by this geodesic and $\Delta t$ the time it took to travel into the bulk and back to the boundary. If we consider planar  holes it is easy to see that boundary causality is violated if $\frac{ \Delta x}{\Delta t} >1$, where $x$ is the transverse coordinate replacing $\phi$ in the planar limit. For spherical black holes, however, the condition has to take into account the compactness of the space, as illustrated in Fig \ref{fig:caus_conditions}. 

We find that,  for  $ ( n-1) < \Delta \phi < n \pi$ , where $n=1,2,3,4....$, causality is violated if 
\begin{align}\label{eq:caus_cond_smbh}
&\frac{\Delta\tilde{\phi}}{\Delta t}=\frac{ (1+ (-1)^{n} (2 n -1)  )\frac{ \pi }{2} -(-1)^{n}   \Delta \phi}{\Delta t} > 1. 
\end{align} 
Our main goal is to determine the region of the parameter space $(r_H, \lambda)$ in which boundary causality is violated, i.e. the region 
for which there exist null geodesics of the appropriate effective metric for which \eqref{eq:caus_cond_smbh} holds. 
In order to do so, it is useful to recall the results from \cite{Buchel:2009tt} (see also appendix \ref{sec:app1}), which found that for planar black holes the range of $\lambda$ in 
which boundary causality is respected is $-7/36 < \lambda < 9/100$. This suggests that for black holes of relative large size (say, $r_H \gtrsim 1$), there will be a similar range of values of $\lambda$ for which there are no geodesics satisfying \eqref{eq:caus_cond_smbh}. 
%\succeq
%
As we decrease $r_H$, we expect this qualitative picture to persist, with probably different values of $\lambda$ characterizing 
the end points of the interval. This turns out to be the case, as we shall now describe. 

\begin{figure}[ht]
\centering
  \includegraphics[width=\textwidth]{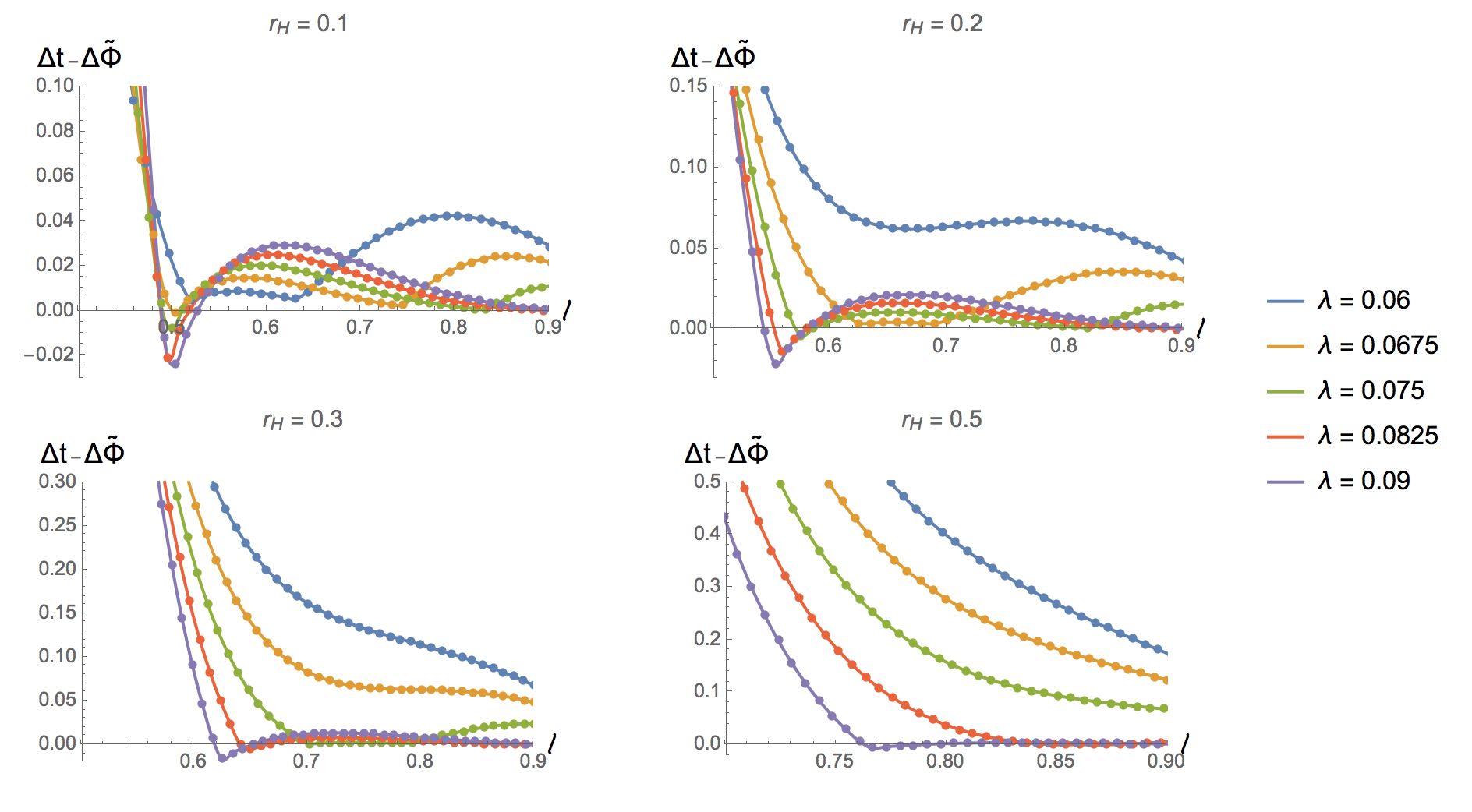}
  \caption{Plot of $(\Delta t - \Delta\tilde{\Phi})$ as a function of $\ell$ for the tensor effective metric with fixed $(r_H, \lambda)$
  with $\lambda >0$. Negative values of $(\Delta t - \Delta\tilde{\Phi})$ indicate violations of boundary causality. 
  The data points are obtained numerically, and we include a polynomial interpolation of the data (solid lines) to guide the eye. 
  As we approach the critical values $\lambda_c^+(r_H)$, the curves start to develop kinks as a result of the definition \eqref{eq:caus_cond_smbh}.} \label{fig:causality_violation_ell}
\end{figure}

\begin{figure}[h!]
\centering
  \includegraphics[width=\textwidth]{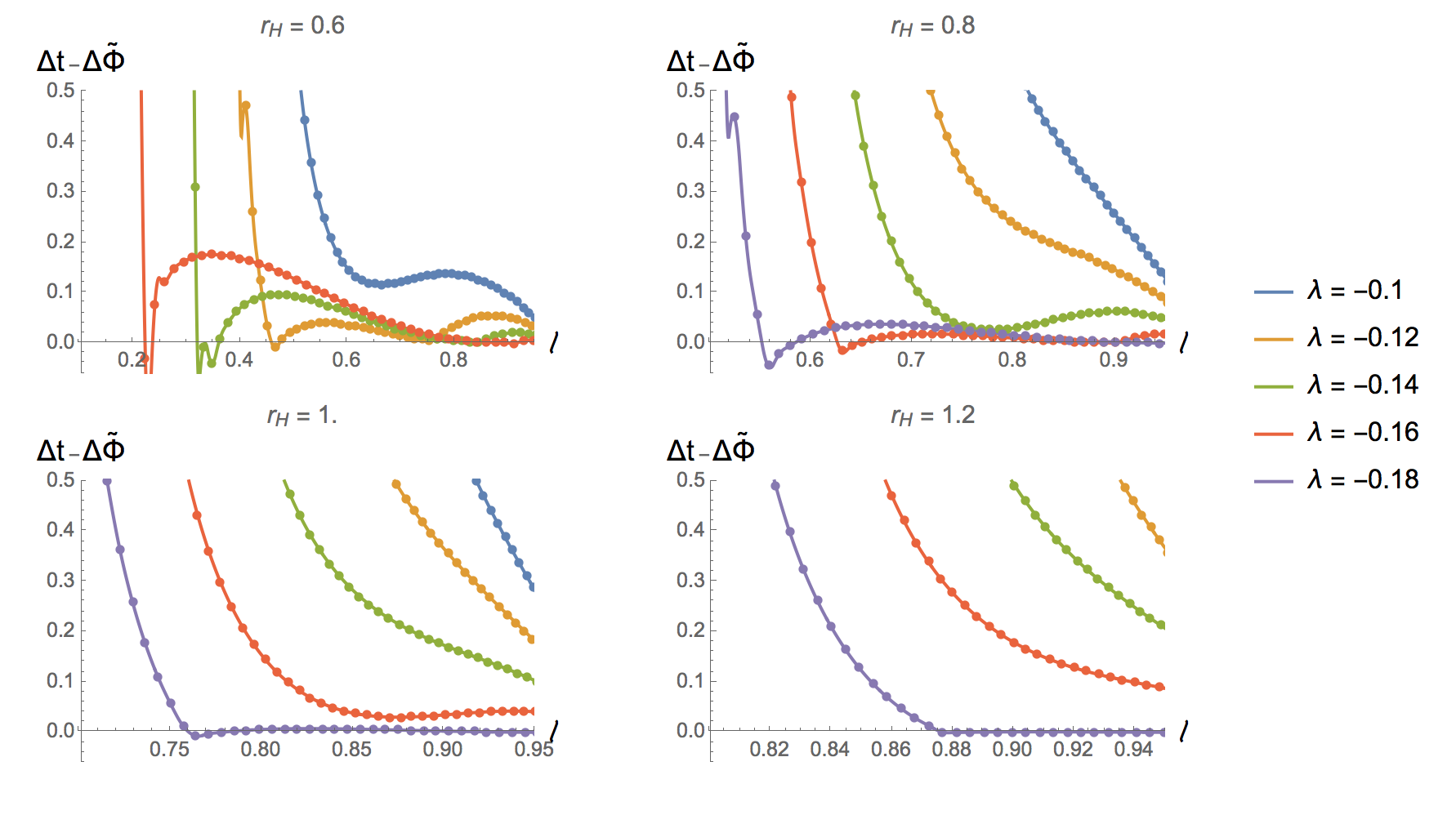}
  \caption{Plot of $(\Delta t - \Delta\tilde{\Phi})$ as a function of $\ell$ for the scalar effective metric with fixed $(r_H, \lambda)$
  with $\lambda < 0$. Negative values of $(\Delta t - \Delta\tilde{\Phi})$ indicate violations of boundary causality. 
  The data points are obtained numerically, and we include a polynomial interpolation of the data (solid lines) to guide the eye. 
  As we approach the critical values $\lambda_c^-(r_H)$, the curves start to develop kinks as a result of the definition \eqref{eq:caus_cond_smbh}.}\label{fig:causality_violation_ell_neg}
\end{figure}

For the sake of concreteness, let us consider $\lambda > 0$ first. In this case, the relevant effective metric is the tensor one. We 
study its null geodesics by fixing $(r_H, \lambda)$ and, using \eqref{eq:delta_t} and \eqref{eq:delta_phi}, we numerically compute 
$\Delta\tilde{\phi}/\Delta t$ in \eqref{eq:caus_cond_smbh}  as a function of $\ell \in (0,1)$. 
See Figs. \ref{fig:causality_violation_ell} and \ref{fig:causality_violation_ell_neg} for these plots with selected values of $(r_H, \lambda)$. 

\begin{figure}[h!]
\center
\subfigure[][]{
\label{fig:summary lpositive}
\includegraphics[width=0.4\linewidth]{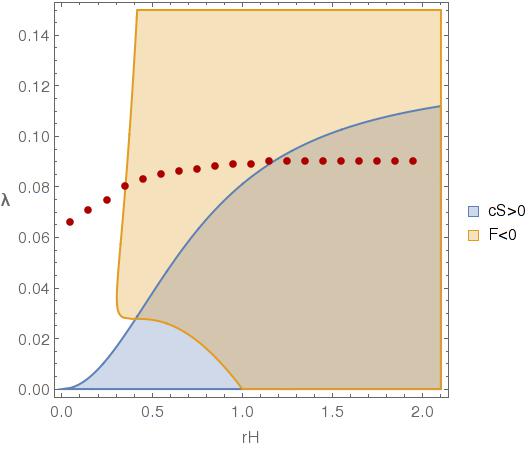}
}\qquad\qquad
\subfigure[][]{
\label{fig:summary lnegative}
\includegraphics[width=0.44\linewidth]{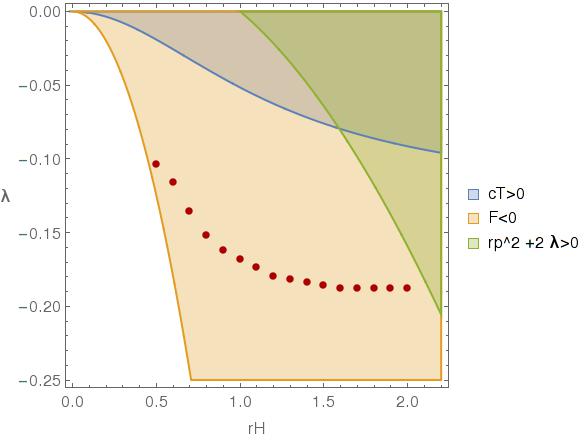}}
\caption{Restrictions on parameter space due to hyperbolicity (blue), thermodynamics (orange), 
the restriction \eqref{restric lambda} (green), 
and boundary causality (red), for $\lambda > 0$ (left) and $\lambda < 0$ (right). Regarding these constraints the allowed region is the
shaded portion of the plot. The set of parameter space allowed by boundary causality comprises 
the region between the red data points and the horizontal axes $\lambda = 0$.   }
\end{figure}

For a given $r_H$, we search for the smallest value of $\lambda$ such that \eqref{eq:caus_cond_smbh} holds for at least one value of $\ell$,%
\footnote{We also observe numerically that this first violating geodesic, occurring in the background with black hole radius $r_H$ and $\lambda=\lambda_c^+$ and for a particular $\ell=\ell_c$, also obeys $\Delta t=\Delta \phi = \pi$.  It turns out there is a simple analytic argument for this, due to S. Fischetti, which we provide in appendix \ref{sec:app2}.}
which we denote by $\lambda_c^+$.
As we can see in Figs. \ref{fig:causality_violation_ell} and \ref{fig:causality_violation_ell_neg}, the derivative of $\Delta\tilde{\phi}/\Delta t$ with respect to $\ell$ grows very large when we approach the critical value $\lambda_c^+$, so in practice we find it convenient to simply construct a grid of values of $\lambda$ with a certain spacing and thus obtain $\lambda_c^+$ with a given resolution (our plots are produced with a  grid of adjustable spacing  $\delta \lambda = 0.0005 - 0.001$)%
\footnote{Had the function $
\Delta\tilde{\phi}/\Delta t$  been better behaved, one could have attempted to find $\lambda_c^+$ by simultaneously solving
the system of equations $\Delta\tilde{\phi}/\Delta t(\ell) = \partial_\ell \Delta\tilde{\phi}/\Delta t(\ell) = 0$ in the variables 
$(\ell, \lambda)$. However, such a method does not perform well in our case.}.
We deal with the negative $\lambda$ case similarly, by taking into account that now the effective metric of interest is the scalar one. 
Following the procedure described above we obtain a lower bound $\lambda_c^-$ below which boundary causality is violated. 
Our method then yields the range $\lambda_c^-(r_H) < \lambda < \lambda_c^+(r_H)$ within which boundary causality is respected. 
We have checked that for black holes of sizes of order one and above our critical values agree with those obtained for planar black holes 
in \cite{Buchel:2009tt}. Our results are summarized in Figs. \ref{fig:summary lpositive} and \ref{fig:summary lnegative}. 

%\begin{figure}[!htm]
%$$
%\begin{array}{ccc}
%  \includegraphics[angle=0,width=0.3\textwidth]{GeoPA.pdf} &\quad \includegraphics[angle=0,width=0.3\textwidth]{GeoPB.pdf} &\quad \includegraphics[angle=0,width=0.3\textwidth]{GeoPC.pdf}
%\end{array}
%$$
%\caption{\small Possible minimal surfaces for region $A \cup B$, in the case of a flat boundary. These correspond to $S_{A \cup B}=2S(\ell)$, $S_{A \cup B}=S(2\ell+x)+S(x)$ and $S_{A \cup B}=2S(\ell+x)$, from left to right, respectively. The latter one will always have a larger area than the other two choices, given that $S(\ell+x)\geq S(\ell)$ for any $x\geq0$.}
%\label{geoP}
%\end{figure}

%Therefore, it is natural to extend the construction of the $r$
%CHI to this scenario by defining it to be given by the CHI of \cite{Hubeny:2012wa} constructed in the geometry 
%\eqref{ds2 GB} for $c_A(r) = c_T(r)$.
%
%In the reminder of this paper we will study such construction in the parameter space spanned by $(\lambda, r_H)$, 
%and present the results for $c_A = c_S$ and $c_A = 1$ -- corresponding to the AdS GB black hole --  for comparison.  
%
%We specifically make a comparison between the allowed combinations of black hole size and Gauss-Bonnet parameter $\lambda$ under two different constraints: the hyperbolicity constraint explored in \cite{Reall:2014pwa}, and the geodesic boundary causality constraint.  We show that hyperbolicity is a stronger constraint for small black holes, while boundary causality is a stronger constraint for large ones.

%%%%%%%%%%%%%%%%%%%%%%%%%%%%%%%%%%%%%%%%%%%%%%%%%%%%%%%%%%%%%%%%%%%%%%%%%%%%%%%%%%%%% 
\section{Results and discussion}
\label{sec:discussion}
%%%%%%%%%%%%%%%%%%%%%%%%%%%%%%%%%%%%%%%%%%%%%%%%%%%%%%%%%%%%%%%%%%%%%%%%%%%%%%%%%%%%%

As depicted in Figs. \ref{fig:summary lpositive}, \ref{fig:summary lnegative}, we have found that five-dimensional theories with a GB coupling $\lambda$ larger than $\sim 0.06$ contain spherical black holes in AdS that violate boundary causality.  As we increase $\lambda$ above this bound, black holes of increasing size begin to violate boundary causality, until at $\lambda=9/100$ even planar black holes 
cause violation. 
Although for ($r_H < r_H^* \sim 1.2$) there are causality-violating black holes at smaller values of $\lambda$, these black holes also violate hyperbolicity, so it isn't possible to (stably) produce them from initial data. 
If we assume solutions which violate hyperbolicity should be removed from the theory, then the precision of the numerics in our boundary causality analysis only allows us to restrict to $\lambda \lesssim 9/100$.

For $\lambda<0$ the situation is simpler.  Again, if $\lambda\lesssim - 0.1$, then the theory contains black holes which violate boundary causality.  However black holes of any size (including planar) for $\lambda<0$ that violate boundary causality also violate hyperbolicity, so if we exclude solutions which violate hyperbolicity then we can place no causality restriction for $\lambda<0$.

Thermodynamic stability places additional constraints on the black hole solutions relevant to a physical ensemble. The thermodynamics of GB black holes has been thoroughly studied \cite{Cvetic:2001bk,Cai:2001dz,Cho:2002hq,Brihaye:2008xu}. In \cite{Cai:2001dz, Cho:2002hq, Liu:2008zf} the free energy was calculated using different methods\footnote{In \cite{Cai:2001dz} the free energy, $F$, was derived integrating the first law while in \cite{Cho:2002hq} it was found using background subtraction to regulate  the Euclidean action and in \cite{Liu:2008zf} the divergences were removed using boundary counterterms.} and they yield the same result, 
\begin{equation}
	 F =\frac{Vol_{S^3}}{16 \pi G_N (r_H^2 + 2 \lambda)}\left[-r_H^6+ r_H^4 -18 \lambda r_H^4 + 3 \lambda r_H^2 + 6 \lambda^2\right] \, .
\end{equation}
The geometries \eqref{ds2_GB} are thermodynamically preferred if $F<0$.
We observe in Figs. \ref{fig:summary lpositive}, \ref{fig:summary lnegative} that the restrictions coming from thermodynamics, boundary causality and hyperbolicity do not always coincide.  However, since thermodynamically unstable solutions are still physical, although not preferred, we must still include these solutions when evaluating the sickness of a theory.

Since our original goal was to find which values of the coupling $\lambda$ lead to a consistent GB theory, we must decide how to proceed when a particular solution of the theory violates boundary causality and/or hyperbolicity.  
If a particular solution violates boundary causality but 
\emph{not} hyperbolicity, as is the case for large ($r_H\gtrsim r_H^*$) black holes in theories with $\lambda \gtrsim 9/100$, then we should throw out the theory as a whole.  At the very least, we should not expect theories with $\lambda\gtrsim 9/100$ to have good CFT duals.

However, if all solutions which produce boundary causality violation also violate hyperbolicity, as is the case for $\lambda \lesssim 9/100$, then we instead conclude that the \emph{solutions} should be thrown out.  We are used to excluding nonphysical solutions, for example naked singularities in general relativity.  Since the solutions in question here are not well-posed initial value problems for any possible starting slice, they are not possible to create from initial data; any small perturbation in the starting data would result in no solution to the equations of motion at all.  Physically we can imagine such perturbations coming from quantum fluctuations or even just from imprecision in the classical setup.

Of course this `hyperbolicity censorship hypothesis',  that these solutions are not produced by generic input data, requires more examination.  If it is in fact possible to start with initial data that later produces a hyperbolicity violation, then the theory (or value of $\lambda$) should be excluded completely.  Additionally, our work is done only with the GB term; it is possible that adding a full set of higher curvature corrections could change the picture substantially.

Our results underline the importance of hyperbolicity of higher derivative theories in a holographic context. In the present work we have addressed the relation between hyperbolicity and boundary causality in a specific example, GB. However the importance of hyperbolicity extends, in some cases,  to the definitions of the constructs themselves. 

Some future directions related to our work include:
\begin{itemize}
\item{ \em Entanglement wedge} Among the various geometric constructs in holography, the entanglement wedge plays an important role in bulk reconstruction. It has been shown \cite{Dong:2016eik} that operators in the bulk can be reconstructed as CFT operators in a given boundary region provided they lie in its entanglement wedge.  It would be interesting to investigate the entanglement wedge construction and its properties  in  theories where different polarizations of the graviton propagate with different speeds and some of them  can be superluminal. 

\item{\em Holographic hyperbolicity?}\\
For large black holes and $\lambda>0$ the question of hyperbolicity was irrelevant since for all values of  the coupling allowed by causality the theory was hyperbolic. We have seen that this is not the case for $\lambda<0$\footnote{Interestingly, very different behaviour for positive and negative GB coupling was also found in other contexts \cite{Brigante:2007nu,deBoer:2009pn,Grozdanov:2016vgg,Caceres:2015bkr,Li:2013cja}}.
As we discussed in the introduction, the analysis of \citep{Camanho:2014apa} indicates that GB may be a pathological toy model for any nonzero $\lambda$, modulo the criticism of \cite{Papallo:2015rna}. However, it might still be interesting to explore the interplay between boundary causality and hyperbolicity of the bulk equations of motion in more general higher derivative theories.  Thus, it is natural to ask  what holographic construction can detect  --in the space of theories--  whether a theory is or is not hyperbolic in a given background.

\item{ \em Non-minimally coupled scalars}\\
Hyperbolicity violation is also known to occur in theories with non-minimally coupled scalars like Horndesnki theory. Some particular cases of Horndeski  do not involve higher derivatives and are known to admit  asymptotically AdS black hole solutions. In this simpler scenario it might be possible to use the Gao-Wald theorem to relate boundary causality to properties of the effective metric.

\item{\em More general theories}\\
We have restricted our analysis to the case of Gauss-Bonnet gravity in five space-time dimensions. It is natural to ask if our results are
significantly modified when considering Gauss-Bonnet in higher dimensions, and/or including more general higher curvature theories. 
Of particular interest would be the case of $R^4$ corrections, since they arise in the context of low energy string theory 
\cite{Gubser:1998nz}. 

\end{itemize}
We hope to return some of these questions in the near future.

\acknowledgments
It is a pleasure to thank Sebastian Fischetti, Simon Caron-Huot, Shiraz Minwalla and Niels Obers for enlightening conversations. We thank the Galileo Galilei Institute for Theoretical Physics for  hospitality and the INFN for partial support during the early stages of this work. E. C. is supported by Mexico's
National Council of Science and Technology (CONACYT) grant CB-2014-01-238734 and by  
the National Science Foundation Grant PHY-1316033. 
T. A. was supported by the European Research Council under the European Union's Seventh Framework Programme
(ERC Grant agreement 307955). He also thanks the partial support of the Newton-Picarte Grant 20140053.
C. K. is supported by the European Union's Horizon 2020 research and innovation programme under the Marie Sk\l{}odowska-Curie grant agreement No 656900, and also in part by the Danish National Research Foundation project ``New horizons in particle and condensed matter physics from black holes''.

%%%%%%%%%%%%%%%%%%%%%%
%Appendix
%%%%%%%%%%%%%%%%%%%%%%%

\appendix
\section{An analytic expansion of $\Delta t$ and $\Delta \phi$ at large $r_m$}\label{sec:app1}
In this section we outline an analytic expansion of \eqref{eq:delta_t} and \eqref{eq:delta_phi} at large $r_m$.  This expansion is sufficient to obtain the planar black hole limits, but is not convergent enough to match with our numerical results for small $r_h$.

We begin by replacing the parameter $\ell$ with $r_m$, using \eqref{eq:def_rm}.  One should keep in mind that the minimum radius $r_m$ refers to the largest root of \eqref{eq:def_rm} for a given $\ell$.  Since we will be taking a large $r_m$ limit later this detail won't affect what follows.

Next, we change coordinates from $r$ to $\theta$, defined by
\begin{equation}
\sin \theta = \frac{r_m}{r}.
\end{equation}
When the geodesic leaves the boundary, $r=\infty$, and $\theta=0$. At its minimum radius $r=r_m$, we find $\theta=\pi/2$.  Thus we obtain
\begin{align}
\Delta t &=
2\int_0^{\pi/2} d \theta \frac{r_m\cos\theta \sqrt{f_\infty}}{\sin^2\theta f(r_m/\sin \theta) \sqrt{1-\sin^2\theta \frac{c_A(r_m/\sin \theta)f(r_m/\sin \theta)}{c_A(r_m)f(r_m)}}},
\\
\Delta \phi &=2\int_0^{\pi/2}d \theta \frac{\cos \theta c_A(r_m/\sin\theta)\sqrt{f_\infty}}{\sqrt{c_A(r_m)f(r_m)}\sqrt{1-\sin^2\theta \frac{c_A(r_m/\sin \theta)f(r_m/\sin \theta)}{c_A(r_m)f(r_m)}}}.
\end{align}
Next, we expand the integrands as a series around large $r_m$ (compared to the other scales given by $\lambda$ and $r_H$).  The resulting series of integrals turns out to be doable, but not sufficiently convergent to obtain useful results (hence the numeric approach we take throughout the main body of the paper).

However, it is still possible to obtain the planar limit result.  We do so by computing both $\Delta t$ and $\Delta \phi$ through ${\cal O} (r_m^{-4})$.  We then find 
\begin{align}
\Delta t-\Delta \phi &=
\frac{3\pi\mu\left(
1+\sqrt{1-4\lambda}-4(1+4s)\lambda
\right)}{32 (1-4\lambda)r_m^4} + \mathcal{O}(r_m^{-8}),
\end{align}
where $s=1$ for tensor modes (relevant for $\lambda>0$), and $s=-1$ for scalar modes (relevant for $\lambda<0$).
We indeed see that for $s=1$, $\Delta t-\Delta \phi$ is positive for $\lambda <9/100$, and for $s=-1$, $\Delta t-\Delta \phi$ is positive for $\lambda>-7/36$, as expected from the planar limit in \cite{Buchel:2009tt}.

More terms can be kept in the $r_m$ expansion, but the results are not illuminating and so not provided here.

\section{First violations of boundary causality}\label{sec:app2}
Either by computing the $x_m^0$ term in $\Delta t$ and $\Delta \phi$ via the method in appendix \ref{sec:app1}, or via the numerical procedure followed in the body of the paper, we find that the first violating geodesic (the one for which $\Delta t$ and $\Delta \phi$ saturate the relationship
\eqref{eq:caus_cond_smbh}) satisfies $\Delta t=\Delta \phi=\pi$.

In fact, there is a nice argument for this numerical (and analytic, in the planar limit) observation.  Imagine that the first violating geodesic arrives back at the boundary at a point $p_a$.  Since $p_a$ is the first violator, it must be on the edge of the boundary casual diamond, say at $\Delta t=\Delta \tilde{\phi} = a \pi$, for $a<1$.  Then, all boundary points $q_a$ inside the causal diamond but above $\Delta t = a\pi$ are reachable from $p_a$ by a future directed timelike curve.  Since $p_a$ was reachable from the original shooting point (at $\Delta t=\Delta \tilde{\phi}=0$) via the lightlike geodesic we followed through the bulk, we can get from the original shooting point to any $q_a$ via a combination of null and timelike curves.

Via \cite{Wald:1984rg} Theorem 8.1.2 (applied to the effective metric), this combination of curves must be deformable into a single timelike curve, indicating that the points $q_a$ are timelike connected to the original shooting point.  Thus the path to $p_a$ cannot be the first boundary-causality-violating path.  

Of course if the first violating geodesic arrives at the tip of the boundary causal diamond, at $\Delta t=\Delta \tilde{\phi} = \pi$, then there are no $q_a$.

We thank Sebastian Fischetti for this argument.

\bibliographystyle{JHEP-2}
\bibliography{chi_gb_bibliography}

\end{document}